\documentclass[12pt,preprint]{emulateapj}

\usepackage{natbib}
\usepackage{longtable}
\def\gtorder{\mathrel{\raise.3ex\hbox{$>$}\mkern-14mu
             \lower0.6ex\hbox{$\sim$}}}
\def\ltorder{\mathrel{\raise.3ex\hbox{$<$}\mkern-14mu
             \lower0.6ex\hbox{$\sim$}}}

\newcommand{\newQSOs}{29 }
\newcommand{\plaQSOs}{12 }

\shorttitle{The Magellanic Quasars Survey. I. Doubling the Number of Known AGNs Behind the SMC}
\shortauthors{Koz{\l}owski, Kochanek \& Udalski}

\begin{document}

\title{The Magellanic Quasars Survey. I. 
Doubling the Number of Known AGNs Behind the Small Magellanic Cloud}

\author{Szymon~Koz{\l}owski\altaffilmark{1,2},
Christopher~S.~Kochanek\altaffilmark{2,3},
Andrzej~Udalski\altaffilmark{1}
}

\altaffiltext{1}{Warsaw University Observatory, Al. Ujazdowskie 4, 00-478 Warszawa, Poland}
\altaffiltext{2}{Department of Astronomy, The Ohio State University, 140 West 18th Avenue, Columbus, OH
43210, USA; simkoz@astronomy.ohio-state.edu; ckochanek@astronomy.ohio-state.edu}
\altaffiltext{3}{The Center for Cosmology and Astroparticle Physics, The Ohio State University,
191 West Woodruff Avenue, Columbus, OH 43210, USA}

\begin{abstract}
We report the spectroscopic confirmation of \newQSOs new, \plaQSOs plausible, and 3 previously known 
quasars behind the central $\sim1.5$ deg$^2$ region of the Small Magellanic Cloud.  These were identified
in a single  {\sc 2df/AAOmega} observation on the Anglo-Australian Telescope of $268$ candidates selected primarily
based on their mid-IR colors, along with a smaller number of optically variable sources in 
OGLE-II close to known X-ray sources.  The low detection efficiency was partly expected
from the high surface density of SMC as compared to the LMC targets and the faintness of many
of them (149 with $I>20$~mag). The expected number of $I<20$~mag quasars in the
field is $\simeq 38$, and we found 15 (22 with plausible) or 40\% (60\%).   We did not attempt
to determine the nature of the remaining sources, although several appear to be new
planetary nebulae.  The newly discovered AGNs can be used as reference points for future proper 
motion studies, to study absorption in the interstellar medium, and to study the physics of quasar 
variability with the existing long-term, highly cadenced OGLE light curves.
\end{abstract}


%########################################################################
\section{Introduction}

Active Galactic Nuclei (AGNs) are one of the most important tools in modern cosmology. 
They are intrinsically the brightest continuously luminous sources in the universe, 
which allows their detection to great cosmological distances and their use as probes
of the universe over cosmic time. AGNs are also believed to play a major role in 
galaxy evolution (e.g., \citealt{2004ApJ...613..109H}), acting as means for quenching star formation 
(e.g., \citealt{2006MNRAS.370..645B,2009ApJ...696..891H}) and ``converting'' blue spirals into 
red elliptical galaxies (e.g., \citealt{2005ApJ...620L..79S}).

Although there are over a hundred thousand known AGNs (e.g., \citealt{2010AJ....139.2360S}),
it has proved difficult to identify quasars in dense stellar fields such as the Large and 
Small Magellanic Clouds (LMC and SMC), the Galactic bulge, or other nearby galaxies. For
example, the expected number of quasars brighter than OGLE's $I<20$ magnitude limit
is $\sim25$/deg$^2$ (Figure~13 in \citealt{2006AJ....131.2766R}) while there are 
$\sim150,000$/deg$^2$ stars with $I<20$~mag in the central
regions of the SMC.  As a result, few quasars are known behind the Magellanic
Clouds (57 in the LMC, 28 in the SMC -- within a 5 degree radius)\footnote{This research has made use of the NASA/IPAC 
Extragalactic Database (NED) which is operated by the Jet Propulsion Laboratory, California Institute of Technology, 
under contract with the National Aeronautics and Space Administration.},
and none have been identified behind the Galactic
bulge.  Most of the known quasars were found by investigating sources with
long time scale, non-periodic variability combined with their colors and magnitudes 
(e.g., \citealt{2002AcA....52..241E, 
2003AJ....125.1330D,2003AJ....125....1G, 2003ApJ...591..204S, 2005A&A...442..495D} and recently 
\citealt{2010ApJ...708..927K,2010arXiv1009.2081M,2010ApJ...714.1194S,2010arXiv1008.3143B,2010arXiv1012.2391P,2011arXiv1101.3316K}),
and a smaller number were found as the (variable) optical counterparts to X-ray sources
(\citealt{2002ApJ...569L..15D,2003AJ....125.1330D,2003AJ....126..734D}).

Quasars behind the Magellanic Clouds or the Galactic bulge have several important scientific applications.
First, they are the best sources for fixing the reference frames needed for proper motion studies.
The recent improvements in the proper motions of the Magellanic Clouds (\citealt{2006ApJ...638..772K, 2006ApJ...652.1213K}, 
\citealt{2008AJ....135.1024P}) all relied on {\it Hubble Space Telescope} ({\it HST}) measurements in fields
centered on quasars. The results were surprising, as the Clouds were found to be moving significantly faster
than previous estimates (e.g., \citealt{2002AJ....124.2639V}) and the tangential motion of the SMC differs significantly
from that of the LMC.  This implies that the Clouds may be on their first pericentric passage
and unbound from the Galaxy (\citealt{2007ApJ...668..949B}), and that they may not be bound to
each other (\citealt{2006ApJ...652.1213K}). These results can be
improved not just by longer temporal baselines for the previously known quasars, but also by
better mapping out the contaminating internal motions of the Clouds using additional quasar fields.  
The second familiar reason for finding background quasars is that bright quasars can be used in 
absorption studies of the interstellar medium (e.g., \citealt{1979ApJ...230...49S, 2002ApJS..140..143B}). 
This requires finding the brightest possible quasars behind the Clouds, which is challenging given the
density of contaminating sources and the low density of bright quasars, with only $\sim$0.01, 1 and 25 
quasars/deg$^2$ brighter than $I<16$, 18 and 20 mag, respectively.

The third application, which may be less familiar, is studying the physics of quasar variability.
\cite{2009ApJ...698..895K} showed that quasar light curves can be well modeled as a damped random
walk, a stochastic process with only three parameters: the mean light curve magnitude, a characteristic 
time-scale and an amplitude.  Their study, using $\sim100$ quasars from the MACHO survey (\citealt{2003AJ....125....1G}) 
and quasars targeted for reverberation mapping (\citealt{1999MNRAS.306..637G,2004ApJ...613..682P}), found
strong correlations between these variability parameters and the physical properties of the quasars
such as black hole mass and luminosity. \cite{2010ApJ...708..927K}, based on $\sim2,700$ quasar candidates 
from \cite{2009ApJ...701..508K} with OGLE-III light curves, further demonstrated that the stochastic
models describe quasar variability well, using a more powerful statistical approach from \cite{1992ApJ...385..404P}
and \cite{1992ApJ...398..169R,1994comp.gas..5004R}.
Recently, \cite{2010ApJ...721.1014M} investigated these correlations in greater detail using the light
curves of $\sim9000$ SDSS quasars, finding that the amplitude of variability on long time scales decreases with 
increasing luminosity and increasing rest-frame wavelength, and that it is also correlated with black hole mass. 
The characteristic time scale for returning to the mean luminosity increases with increasing wavelength and 
increasing black hole mass, but remains constant with redshift and luminosity.  There are discrepancies,
however, between the time scale distributions in \cite{2010ApJ...721.1014M} and  \cite{2010ApJ...708..927K} 
which may be due to higher than estimated stellar contamination in  \cite{2010ApJ...708..927K}, selection
differences or the better light curves available for sources in microlensing fields. 

The individual parameter estimates for the SDSS quasars are significantly worse than those for the Cloud
quasars, and this will also be true of light curves from other modern surveys like 
the Catalina Survey (e.g., \citealt{2003DPS....35.3604L,2009ApJ...696..870D}),
the Palomar Transient Factory (e.g., \citealt{2009PASP..121.1395L}), 
Pan-STARRS (e.g., \citealt{2002SPIE.4836..154K}) or 
LSST (e.g., \citealt{2009arXiv0912.0201L}) 
that attempt to cover large fractions of the sky
rather than localized regions.  The OGLE microlensing survey has monitored the LMC, SMC and the Galactic bulge for 
$\sim15$ years, providing light curves for 400 million objects (\citealt{1997AcA....47..319U,2008AcA....58...69U}). 
These fields have also been observed for shorter periods of time by the EROS (e.g., \citealt{2007A&A...469..387T}), 
MACHO (e.g., \citealt{2000ApJ...542..281A}), MOA (e.g., \citealt{2005MNRAS.356..331S}), and SuperMACHO 
(e.g., \citealt{2005IAUS..225..357B}) microlensing surveys, aiming primarily at detection 
of dark matter compact objects in the Galactic halo (e.g,
\citealt{2000ApJ...542..281A, 2007A&A...469..387T, 2009MNRAS.397.1228W}). 

\begin{figure}
\centering
\includegraphics[width=8cm]{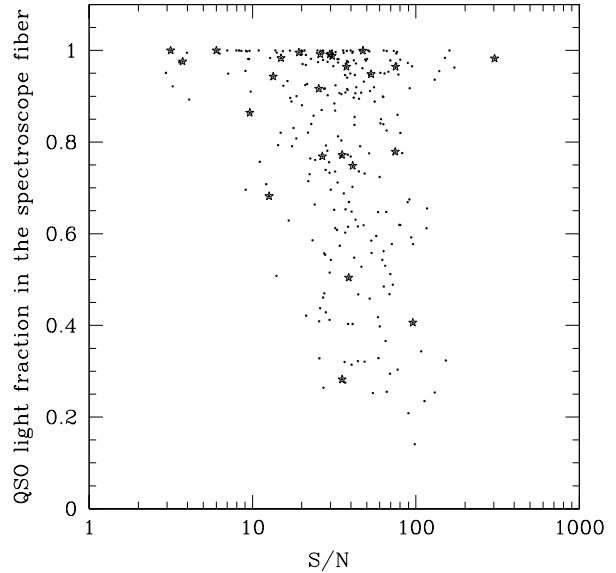}
\caption{Signal-to-noise (S/N) and stellar flux contamination for each of the 268 observed objects. 
The stellar flux contaminating a fiber was the contribution from stars within 10 arcsec of each target assuming a 2 arcsec fiber diameter and
seeing FWHM.  Dots represent all targeted objects, while star symbols are confirmed quasars, and there are no
obvious trends in the confirmed fraction with either S/N or contamination.}
\label{fig:contamination}
\end{figure}

This leaves the problem of identifying the quasars, spectroscopically confirming them, and determining
their properties (luminosities, black hole mass estimates from emission line widths).  \cite{2009ApJ...701..508K}
pointed out that the mid-IR AGN selection method of \cite{2005ApJ...631..163S} would also work reasonably
well in dense stellar fields because the mid-IR colors of quasars are different from all normal stars.
Based on a simple three-level criteria, they selected $\sim5,000$ quasar candidates behind
the Magellanic Clouds from the SAGE (\citealt{2006AJ....132.2268M,2006AJ....132.2034B}) and 
S3MC (\citealt{2007ApJ...655..212B}) data.  In this paper, we present the results of our AAOmega 
spectroscopic follow-up of a single $\sim1.5$ deg$^2$ field in the SMC,
containing 268 quasar candidates from \cite{2009ApJ...701..508K}. We report the confirmation of 
\newQSOs new quasars, and include a list of \plaQSOs plausible quasars.
In Section~\ref{sec:data} we present the data analysis procedures. 
In Section~\ref{sec:newquasars} we present newly confirmed quasars, 
and in Section~\ref{sec:deteff} we discuss the detection efficiency. 
The paper is summarized in Section~\ref{sec:summary}.

%%%%%%%%%%%%%%%%%%%%%%%%%%%%%%%%%%%%%%%%%%%%%%%%%%%%%%%%%%%%%%%%%%%%%%%%%%%%%%%%%%%

\section{Data Analysis}
\label{sec:data}

In \cite{2009ApJ...701..508K} we classified the candidates based on three criteria.  First,
objects in the  \cite{2005ApJ...631..163S} mid-IR color selection region were classed as 
``A'' if they lay away from the track of cool black bodies through the selection region and ``B'' if
they lay close to the track.  Second, the brighter sources have mid-IR magnitudes also
observed for young stellar objects (YSOs), so sources were classified as ``YSO'' if they lay
in the region heavily contaminated by YSOs and as ``QSO'' if they did not.  Finally, objects
with the mid-IR to optical colors typical of quasars were classed as ``a'', while those
which did not were classed as ``b''.   For out first attempt at spectroscopic follow-up,
we selected 249 QSO-Aa and 13 QSO-Ba quasar candidates, for a total of 262 targets.  The
sample included two previously confirmed quasars in the field ($z=0.563$ and $z=1.055$, 
\citealt{2003AJ....125....1G}).  We also searched for optically variable objects in 
OGLE-II\footnote{\tt http://ogledb.astrouw.edu.pl/$\sim$ogle/photdb/} (\citealt{2005AcA....55...43S,1997AcA....47..319U})
within the 3$\sigma$ position uncertainties of the X-ray sources from \cite{2000A&AS..142...41H}.
We targeted 6 X-ray/variability-selected objects, including one at $z=1.06$ confirmed in \cite{2003AJ....125.1330D}.
Since this was an experiment, we did not attempt to exclude known stellar objects from
the target list (see Table~\ref{tab:results2}).  This was one of several fields prepared for the LMC and
SMC, but the only one to be observed.

\begin{figure}
\centering
\includegraphics[width=8cm]{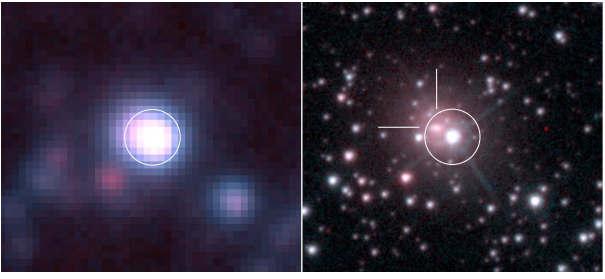}
\caption{QSO J004818.76$-$732059.6 at $z=0.109$, one of our 29 new spectroscopically confirmed quasars, 
as seen by the OGLE 1.3-m ground-based telescope (left) and {\it HST}  (right).  The blue and
red colors are for V/F555W and I/F814W, respectively.  The circle has the 2 arcsec diameter of
the AAOmega fibers, and is also comparable to the FWHM during the observations.  The flux
in the fiber is a blended combination of the bright blue star and the fainter, redder
extended AGN seen in the HST image. 
The image dimensions are 10'' $\times$ 10''. North is up, East is to the left. }
\label{fig:hst}
\end{figure}

The field was centered at (R.A., decl.$)=($00:52:00.0, $-$72:48:00.0) J2000.  While AAOmega has
a field of view of $\sim3$ deg$^2$ (\citealt{2006SPIE.6269E..14S}), the S3MC survey had a 
complex shape so that the actual overlapping area was $\sim1.5$ deg$^2$.  
The spectra were obtained as a service observation on August 17 2009 in moderate sky conditions with
$\sim2$ arcsec seeing.  We used the lowest resolution mode, with $R=1300$ and a spectral coverage
of approximately $5100$\AA\, between the blue (580V) and red (385R) channels. We obtained
three exposures of $1800$~seconds, leading to signal-to-noise ratios (S/N) ranging from 3 to 300,
with 50\% of the objects having $\rm S/N>30$ (Figure~\ref{fig:contamination}). We used
25 sky fibers, positioned to avoid stellar emission based on the OGLE-III SMC
catalogs (\citealt{2008AcA....58..329U}). The
initial {\sc FLD} files were created with the {\sc Configure} software, 
and the final analysis was done using the
{\sc 2dfdr} software (\citealt{1996ASPC..101..195T}).

\begin{figure*}
\centering
\includegraphics[width=17cm]{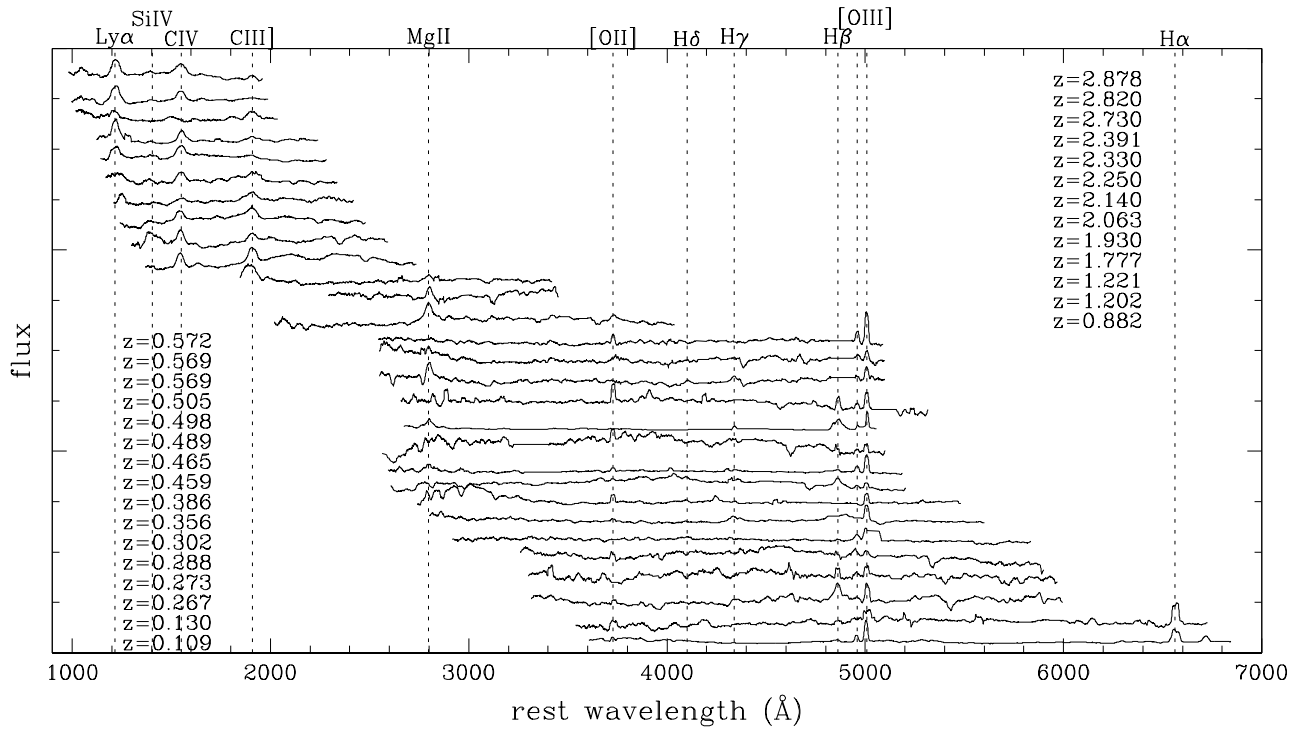}
\caption{Rest-frame spectra of the \newQSOs new quasars behind the SMC, ordered by redshift. 
  The prominent $z=0$ contaminating lines found in almost all the spectra have been masked,
  and we have flattened the continuum in order to focus on the emission lines.
      }
\label{fig:spectra1}
\end{figure*}

The most prominent feature of essentially every spectrum is the emission line contribution of the
interstellar medium (ISM) in the SMC.  These emission lines are redshifted by $\sim2.5$\AA, 
consistent with the radial velocity of the SMC (e.g., \citealt{1987A&A...171...33R}). Since
the stellar densities are high, contamination from nearby stars is also common, as shown by
the OGLE and HST images of the ``brightest'' new AGN in Figure~\ref{fig:hst}.  We estimated the
contribution from nearby stars to the flux in the fiber by computing the contribution from
stars within 10 arcsec to a 2 arcsec diameter fiber assuming a 2 arcsec FWHM Gaussian seeing
profile based on the OGLE-III photometric catalogs (\citealt{2008AcA....58..329U}). 
Figure~\ref{fig:contamination} shows our estimate of this contaminating flux from stars
relative to the flux of the quasar.  Because we are looking for emission line objects,
there was no trivial relationship between either S/N or contamination and our ability
to identify quasars.

%%%%%%%%%%%%%%%%%%%%%%%%%%%%%%%%%%%%%%%%%%%%%%%%%%%%%%%%%%%%%%%%%%%%%%%%%%%%%%%%%%%

\section{New Quasars}
\label{sec:newquasars}

We searched each spectrum for the common (redshifted) quasar emission lines (e.g., \citealt{2001AJ....122..549V})
Ly$\alpha$ at 1216\AA, H$\delta$ at 4101\AA, H$\gamma$ at 4340\AA, H$\beta$ at 4861\AA, H$\alpha$ at 6563\AA, magnesium
MgII at 2800\AA, carbon CIV at 1549\AA~and CIII] at 1909\AA, as well as the 
the narrow forbidden lines of oxygen [O II] at 3727\AA, [O III] at 4959\AA~or 5007\AA.
In general, we required the identification of two lines, except in the redshift range 
$0.7 < z < 1.2$, where we can only find MgII despite the broad spectral coverage.  We identified
\newQSOs new quasars behind the SMC.  Their parameters (including the identified lines) are presented
in Table~\ref{tab:results1} (top panel), and their spectra are shown in Figure~\ref{fig:spectra1}.  We also
confirmed the 3 known quasars we included in the target list.  Another  \plaQSOs were plausibly
quasars with broad emission lines but remained somewhat ambiguous due to the low quality of
their spectra. These objects are listed in Table~\ref{tab:results1} (middle panel).
Table~\ref{tab:results2} lists the remaining sources and any identifications available
from SIMBAD\footnote{This research has made use of the SIMBAD database,
operated at CDS, Strasbourg, France}.  
We have not attempted to classify any of the stellar spectra ourselves.

\section{Detection Efficiency}  
\label{sec:deteff}

The low yield requires some discussion.  \cite{2009ApJ...701..508K} identified 657 quasar candidates 
behind the SMC, and 4699 behind the LMC. Of these, 215 SMC and 1296 LMC candidates are brighter 
than $I<20$ mag, corresponding to 72 and 32 candidates per square degree in the SMC and
LMC, respectively.  The expected number of quasars with $I<20$~mag is $\sim 25$ per
square degree (\citealt{2006AJ....131.2766R}), so we already had indications that the
level of contamination was significantly higher in the SMC/S3MC candidate list than
in the LMC/SAGE candidate list.  For the roughly $1.5$~deg$^2$ overlap of the S3MC
field and the AAOmega field of view, we would expect $38$ quasars with $I<20$~mag. 
Of the observed 268 quasar candidates we were able to confirm 29 new quasars or
11\% of the targets. Of these 29 quasars, 15 are brighter than $I<20$ mag,  which
is 40\% of the expected number (15 out of an expected 38).  

We noted after the observations were complete, that there is an $\sim$0.1~mag blueward 
shift in the $[3.6]-[4.5]$ colors between the $\sim$3 deg$^2$ S3MC data release of 
\cite{2007ApJ...655..212B} used by \cite{2009ApJ...701..508K} and the
subsequent release of 30 deg$^2$ (\citealt{2010AAS...21545921G}). 
This would shift significant numbers of contaminating sources into the mid-IR color
selection region.  We may also have overestimated the purity of the SMC mid-IR selected sample.
This is suggested by the lower density of candidates (40 per deg$^2$) identified in the LMC by 
\cite{2011arXiv1101.3316K} and that the quasar variability time scale distribution
of the mid-IR candidates in \cite{2010ApJ...708..927K} seems to extend to shorter
time scales than those in the SDSS sample considered by  \cite{2010ApJ...721.1014M}.
We will investigate these issues in more detail once we have completed more spectroscopic fields.

%%%%%%%%%%%%%%%%%%%%%%%%%%%%%%%%%%%%%%%%%%%%%%%%%%%%%%%%%%%%%%%%%%%%%%%%%%%%%%%%%%%

\section{Summary}
\label{sec:summary}

With the identification of \newQSOs quasars, we have doubled the number of known quasars behind 
the SMC.  While the yields were lower than expected, there was already reason to expect
this from the difference in the surface density of candidates behind the LMC and SMC
in the candidate catalogs of \cite{2009ApJ...701..508K}.  Fortunately, with the large
numbers of fibers available on AAOmega, there is little problem in targeting the
candidates.  We suspect that the yields in the LMC will be significantly higher.

Identifying large numbers of quasars in the microlensing regions is important not
only for use as probes of the LMC/SMC through proper motions or absorption
studies.  It is also important because of the revolution in studying quasar 
variability that began in 2009 with the introduction of 
a successful method for quantitatively modeling quasar variability as a damped random
walk by \cite{2009ApJ...698..895K}, and its subsequent improvements in \cite{2010ApJ...708..927K} 
and \cite{2010ApJ...721.1014M}.  With the long, well-cadenced OGLE-II, OGLE-III and currently 
OGLE-IV light curves, these new quasars will be invaluable for studying the variability 
properties of quasars if they can be identified in large enough numbers.  Even with
the yields of the current observations, we would expect the full \cite{2009ApJ...701..508K}
candidate list to produce $\sim$1000 new quasars.    

%%%%%%%%%%%%%%%%%%%%%%%%%%%%%%%%%%%%%%%%%%%%%%%%%%%%%%%%%%%%%%%%%%%%%%%%%%%%%%%%%%%

\acknowledgments

We thank Rob Sharp, our service time observer, for his help with the AAOmega data reduction. 
We thank Kris Stanek for helpful comments.
Support for this work was provided by the Polish Ministry of Science and Higher Education
through the program "Iuventus Plus", award number 11-04-00-59-57 to S.K.
This work has been supported by NSF grants AST-0708082 and AST-1009756 to C.S.K and S.K.
S.K. and A.U. are also supported by the European Research Council under 
the European Community's Seventh Framework Programme (FP7/2007-2013),
ERC grant agreement no. 246678 to A.U. 
Based on observations made with the NASA/ESA Hubble Space Telescope, 
obtained from the data archive at the Space Telescope Institute 
(Program ID 10126, PI Olszewski). 
STScI is operated by the association of Universities for Research in Astronomy, 
Inc. under the NASA contract  NAS 5-26555.

%%%%%%%%%%%%%%%%%%%%%%%%%%%%%%%%%%%%%%%%%%%%%%%%%%%%%%%%%%%%%%%%%%%%%%%%%%%%%%%%%%%

%%%%%%%%%%%%%%%%%%%%%%%%%%%%%%%%%%
\clearpage

\begin{deluxetable}{lccccccclr}
\tabletypesize{\tiny}
\tablecaption{Parameters of \newQSOs New, \plaQSOs Plausible and 3 Previously Known Quasars Behind the SMC\label{tab:results1}. }
\tablehead{Name & R.A. & Decl. & V & I & z & OGLE-III & KK09 & Emission & Notes\\
 & & & (mag) & (mag) &  & ID$^a$ & Class & Lines & }
\startdata
\multicolumn{9}{c}{ }\\
\multicolumn{10}{c}{\newQSOs New Spectroscopically Confirmed Quasars}\\
\hline
J003957.65$-$730603.6  &  00:39:57.65  &  $-$73:06:03.60  &  19.74  & 19.44  &  0.569  & 125.5.6063  & ---    & H$\delta\gamma$, [OIII] & X, V  	     \\ %SK\_H286 
J004145.04$-$725435.9  &  00:41:45.04  &  $-$72:54:35.90  &  20.21  & 19.05  &  0.267  & 126.8.16111 & ---    & [OII], H$\beta$, [OIII] & X, V  	     \\ %SK\_H259
J004507.50$-$724121.8  &  00:45:07.50  &  $-$72:41:21.83  &  20.10  & 19.38  &  2.140  & 126.3.17176 & QSO-Aa & CIV, CIII] & ---			     \\ %S3MC$-$AGNc$-$0011
J004551.62$-$725735.0  &  00:45:51.62  &  $-$72:57:34.95  &  20.35  & 19.34  &  2.250  & 126.1.26007 & QSO-Ba & CIV, CIII] & ---			     \\ %S3MC$-$AGNc$-$0405
J004736.12$-$724538.2  &  00:47:36.12  &  $-$72:45:38.19  &  21.77  & 20.15  &  0.572  & 101.8.38227 & QSO-Aa & MgII, [OII], [OIII] & ---		     \\ %S3MC$-$AGNc$-$0064
J004753.62$-$724350.6  &  00:47:53.62  &  $-$72:43:50.55  &  21.21  & 20.66  &  2.730  & 101.8.38968 & QSO-Aa & Ly$\alpha$, CIII] & ---			     \\ %S3MC$-$AGNc$-$0069
J004818.76$-$732059.6  &  00:48:18.76  &  $-$73:20:59.63  &  16.57  & 16.69  &  0.109  & 100.8.45221 & QSO-Aa & [OII], H$\beta$, [OIII], H$\alpha$ & Y, X, B1\\ %S3MC$-$AGNc$-$0078  
J004831.50$-$732339.9  &  00:48:31.50  &  $-$73:23:39.87  &  21.40  & 20.32  &  1.202  & 100.8.19174 & QSO-Aa & CIII], MgII & X 			     \\ %S3MC$-$AGNc$-$0083
J004944.41$-$725400.5  &  00:49:44.41  &  $-$72:54:00.53  &  21.27  & 20.12  &  0.569  & 100.5.55472 & QSO-Aa & MgII, [OIII]& ---			     \\ %S3MC$-$AGNc$-$0106
J005203.27$-$723830.6  &  00:52:03.27  &  $-$72:38:30.58  &  20.80  & 19.90  &  2.391  & 101.2.12689 & QSO-Aa & Ly$\alpha$, SiIV, CIV, CIII] & --- 	     \\ %S3MC$-$AGNc$-$0138
J005235.03$-$732940.6  &  00:52:35.03  &  $-$73:29:40.62  &  19.97  & 19.66  &  2.820  & 103.4.43395 & QSO-Aa & Ly$\alpha$, SiIV, CIV & ---		     \\ %S3MC$-$AGNc$-$0146
J005254.15$-$725628.1  &  00:52:54.15  &  $-$72:56:28.07  &  21.47  & 20.53  &  0.288  & 100.4.26477 & QSO-Aa & [OII], [OIII] & ---			     \\ %S3MC-AGNc-0153
J005259.10$-$724510.9  &  00:52:59.10  &  $-$72:45:10.93  &  20.80  & 20.33  &  1.930  & 101.1.53779 & QSO-Aa & CIV, CIII] & ---			     \\ %S3MC$-$AGNc$-$0156
J005433.79$-$730924.0  &  00:54:33.79  &  $-$73:09:23.97  &  21.61  & 20.37  &  1.221  & 106.5.4994  & QSO-Aa & CIII], MgII & ---			     \\ %S3MC$-$AGNc$-$0177
J005444.70$-$724813.7  &  00:54:44.70  &  $-$72:48:13.67  &  21.62  & 20.73  &  0.505  & 105.7.34076 & QSO-Aa & [OII], H$\beta$, [OIII] & ---		     \\ %S3MC$-$AGNc$-$0178 
J005613.33$-$723821.0  &  00:56:13.33  &  $-$72:38:21.00  &  19.88  & 18.89  &  0.489  & 105.6.41241 & QSO-Aa & [OII], H$\beta$, [OIII] & ---		     \\ %S3MC$-$AGNc$-$0192 
J005708.12$-$722257.8  &  00:57:08.12  &  $-$72:22:57.82  &  21.49  & 20.39  &  0.273  & 108.8.63047 & QSO-Aa & [OII], H$\beta$, [OIII] & ---		     \\ %S3MC$-$AGNc$-$0204
J005714.13$-$723342.8  &  00:57:14.13  &  $-$72:33:42.83  &  20.81  & 20.05  &  0.459  & 105.5.21805 & QSO-Aa & [OII], H$\delta\gamma\beta$, [OIII] & X      \\ %S3MC$-$AGNc$-$0206
J010005.70$-$715723.5  &  01:00:05.70  &  $-$71:57:23.53  &  19.09  & 18.66  &  0.882  & 108.4.10915 & QSO-Aa & MgII, [OII] & X 			     \\ %S3MC$-$AGNc$-$0245
J010046.02$-$722131.4  &  01:00:46.02  &  $-$72:21:31.36  &  20.75  & 20.05  &  0.386  & 108.1.53405 & QSO-Aa & [OII], H$\delta\gamma\beta$, [OIII] & ---     \\ %S3MC-AGNc-0252
J010057.77$-$722230.8  &  01:00:57.77  &  $-$72:22:30.82  &  21.85  & 20.73  &  0.465  & 108.1.55483 & QSO-Aa & MgII, [OII], H$\gamma\beta$, [OIII] & X      \\ %S3MC$-$AGNc$-$0256  
J010127.75$-$721306.2  &  01:01:27.75  &  $-$72:13:06.16  &  20.58  & 19.79  &  2.878  & 113.7.24358 & QSO-Aa & Ly$\alpha$, SiIV, CIV, CIII] & X	     \\ %S3MC$-$AGNc$-$0262 
J010137.52$-$720418.9  &  01:01:37.52  &  $-$72:04:18.94  &  19.31  & 18.51  &  0.356  & 113.6.23341 & QSO-Aa & [OII], H$\delta\gamma\beta$, [OIII] & X, B2\\ %S3MC$-$AGNc$-$0265
J010151.81$-$724110.0  &  01:01:51.81  &  $-$72:41:09.95  &  21.91  & 20.48  &  0.302  & 110.6.2324  & QSO-Aa &  H$\delta\gamma$, [OIII] & ---		     \\ %S3MC$-$AGNc$-$0269
J010241.90$-$723238.7  &  01:02:41.90  &  $-$72:32:38.70  &  19.21  & 18.91  &  0.498  & 110.5.5649  & ---    & H$\delta\gamma\beta$, [OIII] & X, V	     \\ %SK\_H187      
J010314.92$-$724307.5  &  01:03:14.92  &  $-$72:43:07.53  &  19.37  & 18.45  &  2.063  & 110.6.5563  & QSO-Aa & CIV, CIII] & ---		     	\\ %S3MC$-$AGNc$-$0298
J010315.49$-$724221.2  &  01:03:15.49  &  $-$72:42:21.19  &  19.85  & 19.08  &  1.777  & 110.6.5575  & QSO-Aa & CIV, CIII], MgII & ---			     \\ %S3MC$-$AGNc$-$0299
J010425.27$-$724540.4  &  01:04:25.27  &  $-$72:45:40.42  &  21.07  & 20.11  &  0.130  & 110.7.29561 & QSO-Aa & [OII], H$\beta$, [OIII], H$\alpha$ & ---     \\ %S3MC$-$AGNc$-$0312 
J010449.64$-$724850.5  &  01:04:49.64  &  $-$72:48:50.53  &  19.69  & 19.13  &  2.330  & 110.7.31965 & QSO-Aa & Ly$\alpha$, CIV & X  \\ 			%S3MC$-$AGNc$-$0316
\hline
\multicolumn{9}{c}{ }\\
\multicolumn{10}{c}{\plaQSOs Plausible Spectroscopically Confirmed Quasars}\\
\hline
J004520.77$-$730301.9  &  00:45:20.77  &  $-$73:03:01.91  &  21.18  &  20.52  &  1.847  & 125.4.54611 & QSO-Aa & CIV, CIII], MgII & ---		     \\ %S3MC$-$AGNc$-$0020  
J004522.89$-$724541.5  &  00:45:22.89  &  $-$72:45:41.46  &  21.23  &  20.77  &  ?      & 126.2.33358 & QSO-Ba & 2 broad lines & ---		     \\ %S3MC$-$AGNc$-$0404  
J004542.49$-$732317.3  &  00:45:42.49  &  $-$73:23:17.27  &  21.92  &  20.29  &  0.785  & 125.2.55437 & QSO-Aa & MgII, [OII], H$\delta$ & ---	     \\ %S3MC$-$AGNc$-$0028  
J004556.83$-$725636.5  &  00:45:56.83  &  $-$72:56:36.52  &  19.24  &  18.51  &  1.443  & 126.1.23941 & QSO-Aa & CIII], MgII & ---		     \\ %S3MC$-$AGNc$-$0030  
J004958.75$-$725952.2  &  00:49:58.75  &  $-$72:59:52.15  &  20.83  &  20.12  &  2.086  & 100.6.59954 & QSO-Aa & CIV, CIII] & ---		     \\ %S3MC$-$AGNc$-$0114 
J005132.67$-$731411.6  &  00:51:32.67  &  $-$73:14:11.57  &  20.36  &  19.44  &  2.450  & 100.2.11515 & QSO-Aa & Ly$\alpha$, SiIV, CIV, CIII] & ---  \\ %S3MC$-$AGNc$-$0134  
J005454.05$-$723726.2  &  00:54:54.05  &  $-$72:37:26.17  &  20.65  &  19.96  &  1.780  & 105.6.35907 & QSO-Aa & CIII], MgII & ---		     \\ %S3MC$-$AGNc$-$0179  
J005511.81$-$723728.3  &  00:55:11.81  &  $-$72:37:28.33  &  20.59  &  19.43  &  0.804  & 105.6.35056 & QSO-Aa & MgII, [OII], abs. H$\delta$ & --- \\ %S3MC$-$AGNc$-$0180  
J005901.04$-$721330.0  &  00:59:01.04  &  $-$72:13:29.95  &  20.99  &  20.47  &  0.874  & 108.2.42880 & QSO-Aa & MgII, H$\delta$? & ---		    \\ %S3MC$-$AGNc$-$0232  
J010005.10$-$715921.7  &  01:00:05.10  &  $-$71:59:21.65  &  20.52  &  19.95  &  2.716  & 108.4.11919 & QSO-Aa & Ly$\alpha$, CIV, CIII] & X	\\ %S3MC$-$AGNc$-$0244
J010031.06$-$722444.1  &  01:00:31.06  &  $-$72:24:44.14  &  19.98  &  19.07  &  3.170  & 108.1.22073 & QSO-Aa & Ly$\alpha$, SiIV, CIV, CIII]? & ---    \\ %S3MC$-$AGNc$-$0247 
J010531.96$-$723721.3  &  01:05:31.96  &  $-$72:37:21.31  &  20.25  &  19.66  &  1.384  & 110.6.32667 & QSO-Aa & CIII], MgII & --- \\ 	       %S3MC$-$AGNc$-$0327
\hline
\multicolumn{9}{c}{ }\\
\multicolumn{10}{c}{3 Previously Known Quasars}\\
\hline
J005534.64$-$722834.4  &  00:55:34.64  &  $-$72:28:34.42  &  18.70  &  18.28  &  0.563  & 105.5.32347 & QSO-Aa & MgII, H$\delta\gamma\beta$, [OIII] &  DG \\ %S3MC$-$AGNc$-$0183
J010128.03$-$724614.1  &  01:01:28.03  &  $-$72:46:14.11  &  19.87  &  19.31  &  1.055  & 105.2.38973 & QSO-Aa & MgII & G \\ %S3MC$-$AGNc$-$0263
J010244.91$-$721521.9  &  01:02:44.91  &  $-$72:15:21.90  &  18.90  &  18.34  &  1.060  & 113.7.6604  & ---    & MgII & X, D%SK\_H123
\enddata
\tablecomments{$^a$the OGLE-III ID format is smcXXX.Y.ZZZZZZ, where XXX is the field number, 
Y is the chip number, and ZZZZZZ is the source number. In the Notes column, 
X is a quasar with the X-ray counterpart (\citealt{2000A&AS..142...41H}), 
B1(B2)-- sources incorrectly classified as high mass 
X-ray binaries in \cite{2005MNRAS.362..879S} (\citealt{2004A&A...414..667H}), 
Y--sources incorrectly classified as YSOs in \cite{2008AJ....136...18W}, and
V--sources variable in OGLE-II, D--quasar discovered by \cite{2003AJ....125.1330D}, and
G--quasar discovered by \cite{2003AJ....125....1G}.}
\end{deluxetable}

%%%%%%%%%%%%%%%%%%%%%%%%%%%%%%%%%%

\begin{deluxetable*}{lcccclcr}
\tabletypesize{\scriptsize}
\tablecaption{Parameters of 224 Remaining Objects With Spectra\label{tab:results2}.}
\tablehead{Name & RA & Dec & V (mag) & I (mag) & OGLE-III ID & KK09 Class &  Notes}
\startdata
J004413.65$-$724302.23 & 00:44:13.65 & $-$72:43:02.23 & 21.418 & 20.233 &  smc126.3.9008 & QSO-Aa &      Radio \\
J004520.57$-$732410.29 & 00:45:20.57 & $-$73:24:10.29 & 17.494 & 19.155 & smc125.2.22473 & QSO-Aa &         PN \\
J004648.88$-$732258.26 & 00:46:48.88 & $-$73:22:58.26 & 21.235 & 20.501 &  smc100.8.4998 & QSO-Aa &        YSO \\
J005808.62$-$720339.19 & 00:58:08.62 & $-$72:03:39.19 & 21.394 & 21.044 & smc108.3.31448 & QSO-Aa &      X-ray \\
J005820.93$-$721832.84 & 00:58:20.93 & $-$72:18:32.84 & 21.314 & 20.989 & smc108.1.35673 & QSO-Aa &  EclipsBin \\
J005732.27$-$724745.40 & 00:57:32.27 & $-$72:47:45.40 & 21.747 & 20.734 & smc105.7.50832 & QSO-Ba & ---none---
\enddata
\tablecomments{The column Notes includes matches to Simbad objects within 3 arcsec radius, 
where Radio--is a radio source, PN--is a planetary nebula,
YSO--is a young stellar object, X-ray--is an X-ray source, 
EclipsBin--is an eclipsing binary and if there is no information
on the nature of an object it was marked with ``---none---''.
(This table is available in its entirety in a machine-readable form in
the online journal. A portion is shown here for guidance regarding its
form and content.)
}
\end{deluxetable*}

%%%%%%%%%%%%%%%%%%%%%%%%%%%%%%%%%%


\begin{thebibliography}{}

\bibitem[Alcock et al.(2000)]{2000ApJ...542..281A} Alcock, C., et al. 2000, \apj, 542, 281u, Y. 2004, \mnras, 352, 233

% absorption lines in QSO  by interstellar medium
\bibitem[Bechtold et al.(2002)]{2002ApJS..140..143B} Bechtold, J., 
Dobrzycki, A., Wilden, B., Morita, M., Scott, J., Dobrzycka, D., Tran, 
K.-V., \& Aldcroft, T.~L.\ 2002, \apjs, 140, 143 

\bibitem[Becker et al.(2005)]{2005IAUS..225..357B} Becker, A.~C., et al. IAU Symp., 255, 357

%Are the Magellanic Clouds on Their First Passage about the Milky Way?
\bibitem[Besla et al.(2007)]{2007ApJ...668..949B} Besla, G., Kallivayalil, 
N., Hernquist, L., Robertson, B., Cox, T.~J., van der Marel, R.~P., \& Alcock, C.\ 2007, \apj, 668, 949 

%SAGE
\bibitem[Blum et al.(2006)]{2006AJ....132.2034B} Blum, R.~D., et al.\ 2006, \aj, 132, 2034 

%S3MC
\bibitem[Bolatto et al.(2007)]{2007ApJ...655..212B} Bolatto, A.~D., et al.\ 2007, \apj, 655, 212 

% AGN Quenching
\bibitem[Bower et al.(2006)]{2006MNRAS.370..645B} Bower, R.~G., Benson, 
A.~J., Malbon, R., Helly, J.~C., Frenk, C.~S., Baugh, C.~M., Cole, S., 
\& Lacey, C.~G.\ 2006, \mnras, 370, 645 

% QSO selection
\bibitem[Butler 
\& Bloom(2010)]{2010arXiv1008.3143B} Butler, N.~R., \& Bloom, J.~S.\ 2010, arXiv:1008.3143 

\bibitem[Dobrzycki et al.(2002)]{2002ApJ...569L..15D} Dobrzycki, A., Groot, P.~J., Macri, L.~M., \& Stanek, K.~Z.\ 2002, \apjl, 569, L15 
\bibitem[Dobrzycki et al.(2003a)]{2003AJ....125.1330D} Dobrzycki, A., Macri, L.~M., Stanek, K.~Z., \& Groot, P.~J.\ 2003a, \aj, 125, 1330
\bibitem[Dobrzycki et al.(2003b)]{2003AJ....126..734D} Dobrzycki, A., Stanek, K.~Z., Macri, L.~M., \& Groot, P.~J.\ 2003b, \aj, 126, 734 
\bibitem[Dobrzycki et al.(2005)]{2005A&A...442..495D} Dobrzycki, A., Eyer, L., Stanek, K.~Z., \& Macri, L.~M. 2005, \aap, 442, 495

% Catalina CRTS Survey
\bibitem[Drake et al.(2009)]{2009ApJ...696..870D} Drake, A.~J., et al.\ 
2009, \apj, 696, 870 

\bibitem[Eyer(2002)]{2002AcA....52..241E} Eyer, L. 2002, Acta Astron., 52, 241
\bibitem[Geha et al.(2003)]{2003AJ....125....1G} Geha, M., et al. 2003, \aj, 125, 1
\bibitem[Giveon et al.(1999)]{1999MNRAS.306..637G} Giveon, U., Maoz, D., Kaspi, S., Netzer, H., Smith, P.~S. 1999, \mnras,306,637

% S3MC
\bibitem[Gordon 
\& SAGE-SMC Spitzer Legacy Team(2010)]{2010AAS...21545921G} Gordon, K.~D., \& SAGE-SMC Spitzer Legacy Team 2010, American Astronomical Society Meeting Abstracts, 215, \#459.21 

% A ROSAT PSPC catalogue of X-ray sources in the SMC region
\bibitem[Haberl et al.(2000)]{2000A&AS..142...41H}
Haberl, F., Filipovic, M.~D., Pietsch, W., Kahabka, P., 2000, A\&AS, 142, 41

% X-ray binaries in SMC
\bibitem[Haberl 
\& Pietsch(2004)]{2004A&A...414..667H} Haberl, F., \& Pietsch, W.\ 2004, \aap, 414, 667 

% AGN and bulges evolution
\bibitem[Heckman et al.(2004)]{2004ApJ...613..109H} Heckman, T.~M., 
Kauffmann, G., Brinchmann, J., Charlot, S., Tremonti, C., 
\& White, S.~D.~M.\ 2004, \apj, 613, 109 

\bibitem[Hickox et al.(2009)]{2009ApJ...696..891H} Hickox, R.~C., et al.\ 
2009, \apj, 696, 891 

% Pan-STARRS
\bibitem[Kaiser et al.(2002)]{2002SPIE.4836..154K} Kaiser, N., et al.\ 
2002, \procspie, 4836, 154 

% The Proper Motion of the Large Magellanic Cloud Using HST
\bibitem[Kallivayalil et al.(2006a)]{2006ApJ...638..772K}
Kallivayalil, N., et al., 2006, \apj, 638, 772

% Is the SMC Bound to the LMC? The Hubble Space Telescope Proper Motion of the SMC
\bibitem[Kallivayalil et al.(2006b)]{2006ApJ...652.1213K}
Kallivayalil, N., van der Marel, R.~P., \& Alcock, C., 2006, \apj, 652, 1213

% Stochastic modeling of AGN variability
\bibitem[Kelly et al.(2009)]{2009ApJ...698..895K} Kelly, B.~C., Bechtold, J., \& Siemiginowska, A. 2009, \apj, 698, 895

% MACHO var selected qsos
\bibitem[Kim et al.(2011)]{2011arXiv1101.3316K} Kim, D.-W., Protopapas, P., 
Byun, Y.-I., Alcock, C., \& Khardon, R.\ 2011, arXiv:1101.3316 

% Mid-IR-selected QSOs behind Magellanic Clouds
\bibitem[Koz{\l}owski \& Kochanek(2009)]{2009ApJ...701..508K} Koz{\l}owski, S., \& Kochanek, C.~S. 2009, ApJ, 701, 508

% Optical AGN variability
\bibitem[Koz{\l}owski et al.(2010)]{2010ApJ...708..927K} Koz{\l}owski, S., et al.\ 2010, \apj, 708, 927 

% CSS
\bibitem[Larson et al.(2003)]{2003DPS....35.3604L} Larson, S., Beshore, E., 
Hill, R., Christensen, E., McLean, D., Kolar, S., McNaught, R., 
\& Garradd, G.\ 2003, Bulletin of the American Astronomical Society, 35, 982 

% PTF
\bibitem[Law et al.(2009)]{2009PASP..121.1395L} Law, N.~M., et al.\ 2009, 
\pasp, 121, 1395 

% LSST
\bibitem[LSST Science Collaboration et al.(2009)]{2009arXiv0912.0201L} 
LSST Science Collaboration, et al.\ 2009, arXiv:0912.0201 

% SDSS QSO variability
\bibitem[MacLeod et al.(2010a)]{2010ApJ...721.1014M} MacLeod, C.~L., et al.\ 
2010, \apj, 721, 1014 

% SDSS quasar selection
\bibitem[MacLeod et al.(2010b)]{2010arXiv1009.2081M} MacLeod, C.~L., et al.\ 2010b, arXiv:1009.2081 

% SAGE
\bibitem[Meixner et al.(2006)]{2006AJ....132.2268M} Meixner, M., et al.\ 2006, \aj, 132, 2268 

% QSO selection
\bibitem[Palanque-Delabrouille et al.(2010)]{2010arXiv1012.2391P} 
Palanque-Delabrouille, N., et al.\ 2010, arXiv:1012.2391 

% Reverberation mapping QSOs
\bibitem[Peterson et al.(2004)]{2004ApJ...613..682P} Peterson, B.~M., et al. 2004, \apj, 613, 682

% Proper Motions of the Large Magellanic Cloud and Small Magellanic Cloud: Re-Analysis of Hubble Space Telescope Data
\bibitem[Piatek et al.(2008)]{2008AJ....135.1024P} Piatek, S., Pryor, C., \& Olszewski, E.~W., 2008, \aj, 135, 1024

\bibitem[Press, Rybicki \& Hewitt(1992)]{1992ApJ...385..404P} Press, W.~H., Rybicki, G.~B., \& Hewitt, J.~N.  1992, \apj, 385, 404 

% SDSS QSOs
\bibitem[Richards et al.(2006)]{2006AJ....131.2766R} Richards, G.~T., et al.\ 2006, \aj, 131, 2766

% SMC radial velocity
\bibitem[Richter et al.(1987)]{1987A&A...171...33R} Richter, O.-G., Tammann, G.~A., \& Huchtmeier, W.~K.\ 1987, \aap, 171, 33 

\bibitem[Rybicki \& Press(1992)]{1992ApJ...398..169R} Rybicki, G.~B., \& Press, W.~H. 1992, \apj, 398, 169 
\bibitem[Rybicki \& Press(1994)]{1994comp.gas..5004R} Rybicki, G.~B., \& Press, W.~H. 1994, Computer, 5004

% Absorption by ISM in QSOs 
\bibitem[Sargent et al.(1979)]{1979ApJ...230...49S} Sargent, W.~L.~W., 
Young, P.~J., Boksenberg, A., Carswell, R.~F., 
\& Whelan, J.~A.~J.\ 1979, \apj, 230, 49

% QSO variability selection
\bibitem[Schmidt et al.(2010)]{2010ApJ...714.1194S} Schmidt, K.~B., 
Marshall, P.~J., Rix, H.-W., Jester, S., Hennawi, J.~F., 
\& Dobler, G.\ 2010, \apj, 714, 1194 

% SDSS QSO catalog
\bibitem[Schneider et al.(2010)]{2010AJ....139.2360S} Schneider, D.~P., et 
al.\ 2010, \aj, 139, 2360 

% AAOmega
\bibitem[Sharp et al.(2006)]{2006SPIE.6269E..14S} Sharp, R., et al.\ 2006, \procspie, 6269,  

% 
\bibitem[Shtykovskiy 
\& Gilfanov(2005)]{2005MNRAS.362..879S} Shtykovskiy, P., \& Gilfanov, M.\ 2005, \mnras, 362, 879 

% Red Ellipticals
\bibitem[Springel et al.(2005)]{2005ApJ...620L..79S} Springel, V., Di 
Matteo, T., \& Hernquist, L.\ 2005, \apjl, 620, L79 

\bibitem[Stern et al.(2005)]{2005ApJ...631..163S} Stern, D., et al. 2005, \apj, 631, 163
\bibitem[Sumi et al.(2003)]{2003ApJ...591..204S} Sumi, T., et al. 2003, \apj, 591, 43
\bibitem[Sumi et al.(2005)]{2005MNRAS.356..331S} Sumi, T., et al. 2005, \mnras, 356, 331

% OGLE-2 DB
\bibitem[Szyma{\'n}ski(2005)]{2005AcA....55...43S} Szyma{\'n}ski, M.~K.\ 2005, Acta 
Astronomica, 55, 43 

%2DFDR
\bibitem[Taylor et al.(1996)]{1996ASPC..101..195T} Taylor, K., Bailey, J., 
Wilkins, T., Shortridge, K., 
\& Glazebrook, K.\ 1996, Astronomical Data Analysis Software and Systems V, 101, 195 

% EROS microlensing towards LMC
\bibitem[Tisserand et al.(2007)]{2007A&A...469..387T} Tisserand, P., et al.\ 2007, \aap, 469, 387 

\bibitem[Udalski, Kubiak \& Szyma{\'n}ski(1997)]{1997AcA....47..319U} Udalski, A., Kubiak, M., \& Szyma{\'n}ski, M. K. 1997, Acta Astron., 47, 319

\bibitem[Udalski et al.(2008a)]{2008AcA....58...69U} Udalski, A., Szyma{\'n}ski, M. K., Soszy{\'n}ski, I., \& Poleski, R. 2008, Acta Astron., 58, 69

% SMC maps
\bibitem[Udalski et al.(2008b)]{2008AcA....58..329U} Udalski, A., et al.\ 2008, Acta Astron., 58, 329 

% SDSS quasar spectrum
\bibitem[Vanden Berk et al.(2001)]{2001AJ....122..549V} Vanden Berk, D.~E., et al.\ 2001, \aj, 122, 549 

% LMC/SMC prop motions
\bibitem[van der Marel et al.(2002)]{2002AJ....124.2639V} van der Marel, 
R.~P., Alves, D.~R., Hardy, E., \& Suntzeff, N.~B.\ 2002, \aj, 124, 2639 

% SDSS quasar variability
\bibitem[Whitney et al. (2008)]{2008AJ....136...18W} Whitney, B.~A., et al., 2008, \aj, 136, 18
\bibitem[Wyrzykowski et al.(2009)]{2009MNRAS.397.1228W} Wyrzykowski, {\L}., 
et al.\ 2009, \mnras, 397, 1228 

\end{thebibliography}
\end{document}